\documentclass[pre,aps,10pt,twocolumn,superscriptaddress,notitlepage,nofootinbib]{revtex4}
\usepackage{epsfig,amsmath,amssymb,amsfonts,graphicx,color,calc,epstopdf}

\let\a=\alpha \let\b=\beta  \let\d=\delta

\let\s=\sigma   
   \let\G=\Gamma
\let\D=\Delta   
   
  \let\th=\theta \let\io=\infty

\newcommand{\beq}{\begin{equation}} 
\newcommand{\eeq}{\end{equation}}
\newcommand{\ba}{\begin{eqnarray}}
\newcommand{\ea}{\end{eqnarray}}

\begin{document}

\title{Jamming transition of randomly pinned systems}

\author{Carolina Brito}
\affiliation{Inst. de Fisica, Universidade Federal do Rio Grande do Sul, CP 15051, 91501-970, Porto Alegre RS, Brazil}

\author{Giorgio Parisi}
\affiliation{Dipartimento di Fisica,
Sapienza Universit\`a di Roma,
INFN, Sezione di Roma I, IPFC -- CNR,
P.le A. Moro 2, I-00185 Roma, Italy
}

\author{Francesco Zamponi}
\affiliation{LPT,
Ecole Normale Sup\'erieure, CNRS UMR 8549, 24 Rue Lhomond, 75005 France}

\begin{abstract}
We consider a system of hard spheres close to jamming, where translation invariance is broken by pinning a randomly chosen set of particles.
Using two different protocols, we generate two kinds of packings at the jamming point, isostatic and hyperstatic packings.
In the case where the packings are isostatic, jamming transition is not affected by random pinning: 
the jamming density is only slightly reduced, a generalized isostaticity condition holds, the system is marginally rigid over the entire
 glass phase,  and the typical structural signatures of jamming are unchanged.  
Besides, random pinning does not have effect on the vibrational modes of the amorphous system, at least on the time and length scale
that we are able to probe.
For packings that are hyperstatic at jamming, the microscopic properties of the system are strongly affected by the random pinning: 
the frozen degrees of freedom  introduce an excess of constraints that drives the system far from the mechanical marginality conditions 
and stabilize the low frequency modes of the system, shifting them to higher frequencies.
The distance from the isostatic point seems to be the only relevant parameter in the system in both cases.
These two cases are in contrast with the behavior of the plane waves of a crystal:
 the whole spectrum in this case is shifted to higher frequencies as soon as some particles are pinned. 
\end{abstract}

\maketitle

\section{Introduction}
%\paragraph*{Introduction --}
At low enough temperature,
all the information on linear elasticity of a system is contained in its vibrational modes. 
In a continuous isotropic elastic medium, translation invariance implies that the vibrational modes are plane
waves and the density of vibrational modes $D(\omega)$ follows the Debye law $D(\omega ) \sim \omega ^{d-1}$, 
where $d$ is the space dimension~\cite{Ashcroft-Mermin}. 
By contrast, amorphous materials exhibit  an excess of low-frequency vibrational modes compared to the Debye behavior  \cite{Phillips81,Sokolov95}, 
often called Boson peak. 
There is no consensus about the origin and nature of these modes in more realistic systems \cite{Ruocco08,Grigera_nature2003,Schirmacher98,Wyart2010},
but some progress has been made for a class of simple amorphous materials, made either by hard spheres or by 
elastic particles that only interact through a finite-range repulsive potential. 
These systems present a well-defined {\it jamming} transition at which the pressure becomes infinite (for hard spheres),
or the overlaps between particles vanish (for elastic particles).
It has been shown that the distance to this point governs many properties of the system: 
scaling laws characterize the microscopic structure~\cite{PhysRevE.68.011306, wyart_thesis,Jacquin2011}, 
elastic~\cite{PhysRevE.68.011306, wyart_thesis, Ellenbroek06} and transport~\cite{Xu-PRL2007} properties, 
and relaxation in shear flows~\cite{Olsson-PRL2007}. The critical regime can be approached equivalently from the
hard sphere or the elastic sphere side~\cite{IBB13}.
An interesting property of these systems close to the jamming point is the presence of the Boson peak, which has been shown to be
a consequence of the fact that the system lives close to the limit of {\it marginal stability}, as expressed by the {\it isostatic condition},
that the number of mechanical constraints equals the number of degrees of freedom in the system.
The soft modes associated to the peak are characterized by a length scale $l^*$ that diverges near the critical point~\cite{wyart_thesis}.

Besides all this progress, there is still some work to be done to understand fully the property of these modes.
In particular, it is not clear whether the soft modes are related to the translational symmetry 
of the Hamiltonian and the associated low frequency phonon modes~\cite{Grigera_nature2003,Ruocco08}, 
or if they could also exist in a system where translational invariance is explicitly broken.
In this paper, we investigate this issue by considering a system where
translational invariance is broken by ``pinning" a randomly chosen fraction of particles $f$
(see~\cite{OR2012} for a related study).
We use two kinds of configurations to study the glass phase near the jamming transition: isostatic and hyperstatic packings, that can be produced using
different pinning protocols.

We first show that the introduction of pinned particles does not change dramatically the physics of the system if 
the initial packings are isostatic: a jamming transition is still observed and its properties are governed by the distance to the marginally stable
isostatic limit, which is now generalized to take into account the frozen degrees of freedom associated to the pinned particles. 
Moreover, the peculiar structural properties associated to the transition, and in particular the divergence of the pair correlation function close to contact,
are not affected by the pinning. 
Concerning the vibrational properties, it is observed that the soft modes of the system are not strongly affected by the random pinning. 
Next, we show that, as soon as hyperstatic packings are used, the entire glass phase is overconstrained. As a consequence, the vibrational modes of low
frequency are stabilized and remain at finite frequency, which a length scale directly associated to the distance to the generalized isostatic limit.
We contrast this behavior with that of plane waves in a crystal and verify that the consequence of pinning some particles is very different in this case: 
because plane waves are a direct consequence of the translational invariance of the system,  we find that the vibrational spectrum of the crystal is modified in the presence of pinning.
We therefore conclude that, at least on the time and length scales we can probe, the soft modes of jammed packings
are not directly associated with the translational invariance of the system, but rather to its isostaticity, as predicted in~\cite{wyart_thesis}.

\section{Numerical Methods}
\label{methods}
%\paragraph*{Methods --}
We consider a system of $N=4096$ bidisperse hard disks ($d=2$) enclosed in a volume $V$ with periodic boundary conditions.
All particles have equal mass $m$;
half of them have diameter $\sigma_1$, and the others
$\sigma_2 = 1.4\sigma_1$.  The packing fraction is $\phi=\pi (\s_1^2 + \s_2^2) N/(8 V)$. 
We approach the jamming transition from lower density
by using the Lubachevsky-Stillinger \cite{DTS05} algorithm, which is based on
event-driven molecular dynamics~\cite{Tildesley}: 
particles travel in a free flight until they collide elastically. 
They start with tiny diameters $\sigma_j^0$ which are inflated along the time 
as $\sigma_j(t) = \sigma_j^0 (1 + \Gamma t)$, where $\Gamma$ is the inflation rate. 
In the following, dimensional quantities will be plotted for given configurations, 
and we will choose the corresponding $\sigma_1$, $k_B T$ 
and $\sqrt{m\sigma_1^2/k_B T}$ as unit of length, energy and time respectively.

\begin{figure}
\centering
\includegraphics[width = 0.95\columnwidth, clip]{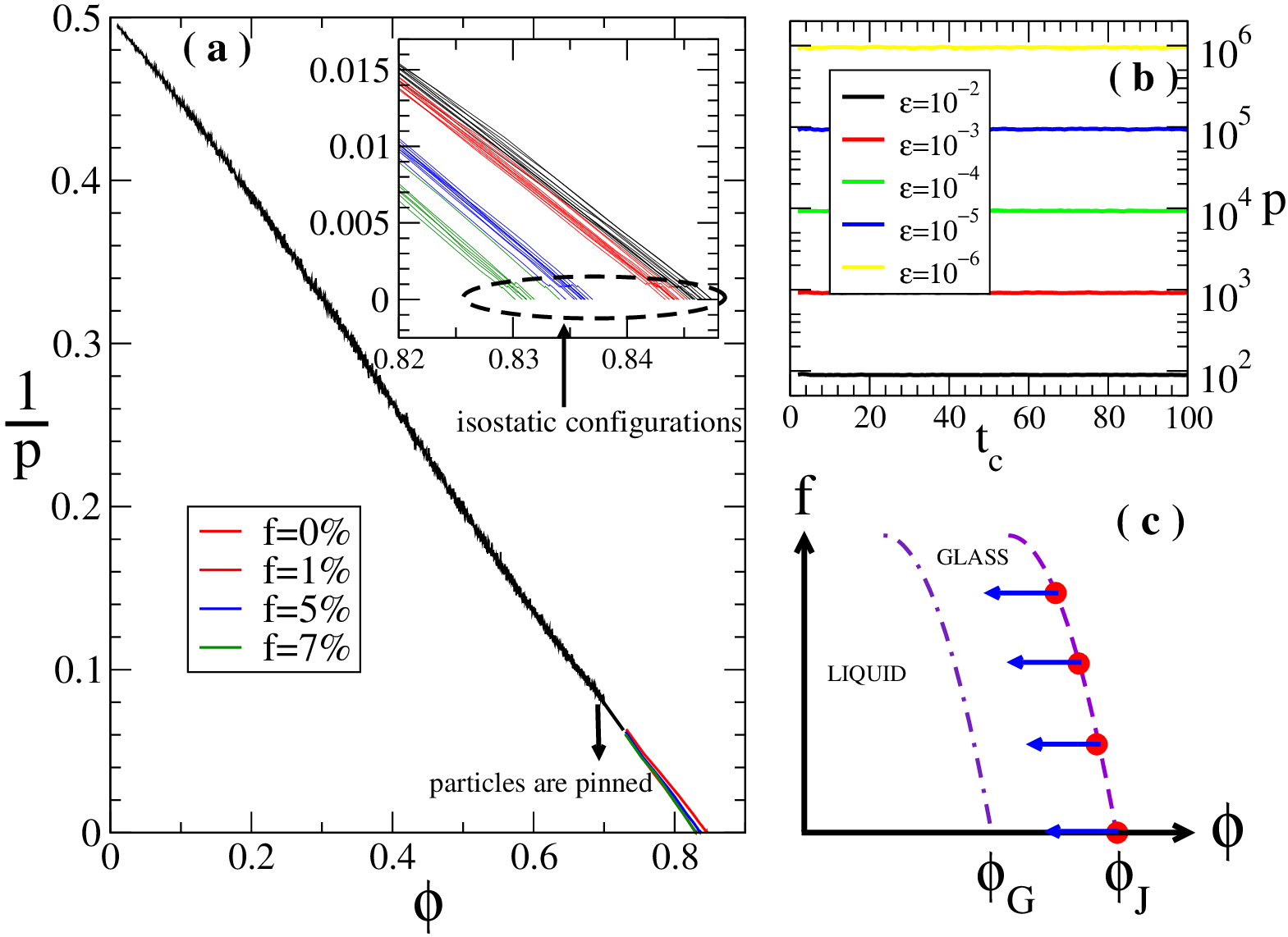}
\caption{Summary of the procedure to generate isostatic configurations with pinned particles and to study its microscopic structure at different pressures. {\bf (a)}: Pressure versus packing fraction for different $f$. Each line correspond to an initial condition.
Inset: zoom in the region $p \sim 10^{12}$. {\bf (b)}: pressure versus computational time for one particular isostatic condition, whose $\s_j$ were reduced to generate configurations at different $p$. {\bf (c)}: schematic phase diagram when a fraction $f$ of particles are pinned. Jamming and glass transitions occur at smaller $\phi$ when $f$ is increased. 
Red circles indicate the location of the isostatic configurations generated by the procedure shown in (a),
and arrows indicate the location of the configurations generated by reducing $\s_j$, as exemplified in (b).
}
\label{scheme}
\end{figure}

\subsection{Protocols to generate isostatic and hyperstatic packings with frozen particles at the jamming point}

%We used to different protocols to generate isostatic and hyperstatic packings.
To generate isostatic packings with pinned particles, we  let the system evolve until it reaches a packing fraction 
$\phi^{(p)} = 0.7$ and we then chose randomly  a set of $N f$  particles and froze their positions. At this point, 
 their diameters $\s_j$ stop increasing, while the rest of the system follows its normal dynamics.
 The reason why we do not inflate the pinned particles is that close-by pinned particle might overlap when inflated.
We varied $\phi^{(p)}$ and verified that the precise choice of this parameter does not affect our results, {\it provided} that the pinning is done 
while the system is still at equilibrium.
Along the simulation, reduced pressure $p = \b V P/N$ and $\phi$ both increase, and 
$\Gamma$ is varied: $\Gamma=10^{-3}$ up to $p=10^{3}$, 
$\Gamma=10^{-4}$ up to $p=10^{5}$ and $\Gamma=10^{-5}$ up to $p=10^{12}$, 
when the simulation stops and the system is considered as jammed at $\phi=\phi_J^f$~\cite{DTS05}.
This procedure is exemplified in Fig.~\ref{scheme}a where $p$ is shown as a function of $\phi$ 
for different values of $f$ and various initial conditions. 
One observes that $\phi_J^f$ decreases on increasing $f$, consistently with the results of Ref.~\cite{OR2012}.
Moreover, the glass transition point $\phi_G^f$ where the curve $p(\phi)$ becomes different from the liquid
equation of state also decreases with increasing $f$~\cite{Cammarota2011}.

We then check if the final configurations at pressure $p=10^{12}$ are indeed isostatic (see below).
Although this is always the case when $f$ is small, 
when $f\geq5\%$ sometimes a subset of the system, surrounded by pinned particles, becomes jammed 
before the whole system is: these configurations are not used in this work.

To generate hyperstatic packings, the procedure is slightly different. 
We do not fix any particle until the system reaches the jamming point, at pressure $p=10^{12}$.
Only at this packing fraction, denoted by  $\phi_J^{f=0}$ and where the system is isostatic~\cite{PhysRevE.68.011306,wyart_thesis}, 
a random set of $N f$  particles has their position and diameters frozen.

\subsection{Protocol to generate packings at smaller pressure}

These fully jammed configurations at $\phi_J^f$, both isostatic and hyperstatic, are used  as initial conditions to study the system at 
different pressures in the glassy phase, as indicated by the arrows in Fig.~\ref{scheme}c.
To do so, we reduce the diameters of all particles by a fraction $\epsilon$, so that
$\phi = \phi_J^f (1-\epsilon)^2$. We assign random velocities to the mobile 
particles, keep the pinned particles fixed, and launch the event-driven simulations with $\Gamma=0$.
To contrast the behavior under pinning of an amorphous system
with that of a crystal, we also considered
a perfectly hexagonal network of monodisperse particles at maximum 
packing fraction $\phi_{cryst}^{max} \approx 0.9$. We pinned $Nf$ randomly chosen particles,
and we repeated the procedure above to obtain pinned crystalline configurations at different pressures.
In the amorphous case, for each $f$ we averaged the data over $\sim 30$ independent isostatic configurations at $\phi_J^f$.
We observe that during the $\G=0$ runs there is no rearrangement of particles in the system,
and the pressure $p$ remains stable (Fig.~\ref{scheme}b), 
indicating that the metastable amorphous glass states we produce have a very long lifetime.

\subsection{Measure of the vibrational modes of the system}

To measure the spectrum of vibrations of the hard sphere system we define a contact force network within an interval 
of time $\Delta t$~\cite{BritoWyart_EPL2006, DTS05, Fergunson2004} which we fix to $\Delta t= 1000N$ collisions.
Two particles are said to be in contact if they collide with each other during $\Delta t$;
we define $h_{ij}$ as the average (over $\Delta t$) gap between two particles and
the contact force $f_{ij}$ as the average momentum they exchange per unit of time.
In a metastable state, it is possible to define an effective potential
$V_{\rm eff} = -k_B T \log h_{ij}$ that allows to compute a dynamical matrix $\cal M$
that describes the linear response of the average displacement of
the particles to an external force~\cite{BritoWyart_EPL2006, BritoWyart_JCP2009}. 
The eingenvectors of $\cal M$ are the normal modes of the system, with frequencies $\omega$ that  
are the square roots of the eigenvalues. 
The distributions of frequencies gives the density of states $D(\omega )$.
Note that the choice $\Delta t= 1000N$ is non trivial, because choosing a smaller $\D t$ would affect
our results. Instead, choosing a larger $\D t$ does not seem to affect the results and we therefore
believe that $\D t =1000 N$ is large enough to obtain a proper averaging of the contact force network.

\section{Results} 

Let us analyze the jammed configurations at $\phi_J^f$. 
The total number of contacts is $N_c = N z/2$ and $z$ is the average connectivity of a particle. 
A global mechanical stability criterion is that the number $N_c$ of independent contact force components
has to exceed the number of degrees of freedom for the packing to be mechanically stable~\cite{PhysRevE.60.687}.
For a system in which particles are pinned while the system is still at equilibirum, 
the latter is $N d (1 - f)$, leading to the generalized isostatic condition $z_J = 2d (1-f)$. 

In the case where particles are pinned at the jamming point, 
the counting in the number of contacts and the number of degrees of freedom goes as following: 
at  $\phi_J^{f=0}$, the packing is isostatic and this implies that  $z_J=2d$ and the 
total  number of contacts in the packing is $Nz/2=Nd$. 
When particles are pinned,
the total number of degrees of freedom is now $Nd(1-f)$, but the number of contacts does not change, 
leading to a situation where 
 the total number of contacts is higher than the degrees of freedom: $ Nz/2 = Nd > Nd(1-f)$, which defines 
 a hyperstatic  packing.

%In the folowing sections we study how the introduction of this lengh scale affects the signature of the jammed packings and the microscopic dynamics during the glassy phase.

\subsection{Structural signatures of jamming in pinned packings} 

To investigate if the pinned particles have an influence on the microscopic structure of the packings at the jamming point, we compute
the integrated pair correlation function $Z(\d)$~\cite{DTS05}, that is defined as follows. For each pair $\langle i j \rangle$ of particles in contact (including
also the pinned particles) we compute the normalized gap 
$\d_{ij} = (2r_{ij} - \s_i - \s_j)/(\s_i + \s_j)$, where $r_{ij}$ is the distance 
between the two particles' centers. Then
\beq
Z(\d) = \frac2N \sum_{\langle i j \rangle} \th(\d_{ij} < \d) 
= \frac1N \sum_{i \neq j} \th(\d_{ij} < \d) 
\ ,
\eeq
where $\langle i j \rangle$ counts the particle pairs, and $\th(x)$ is the Heaviside step function. In other words, $Z(\d)$ counts the average number of particles that lie at normalized distance smaller than $\d$
with respect to a reference particle.

\begin{figure}
\centering
\includegraphics[width = 0.95\columnwidth, clip]{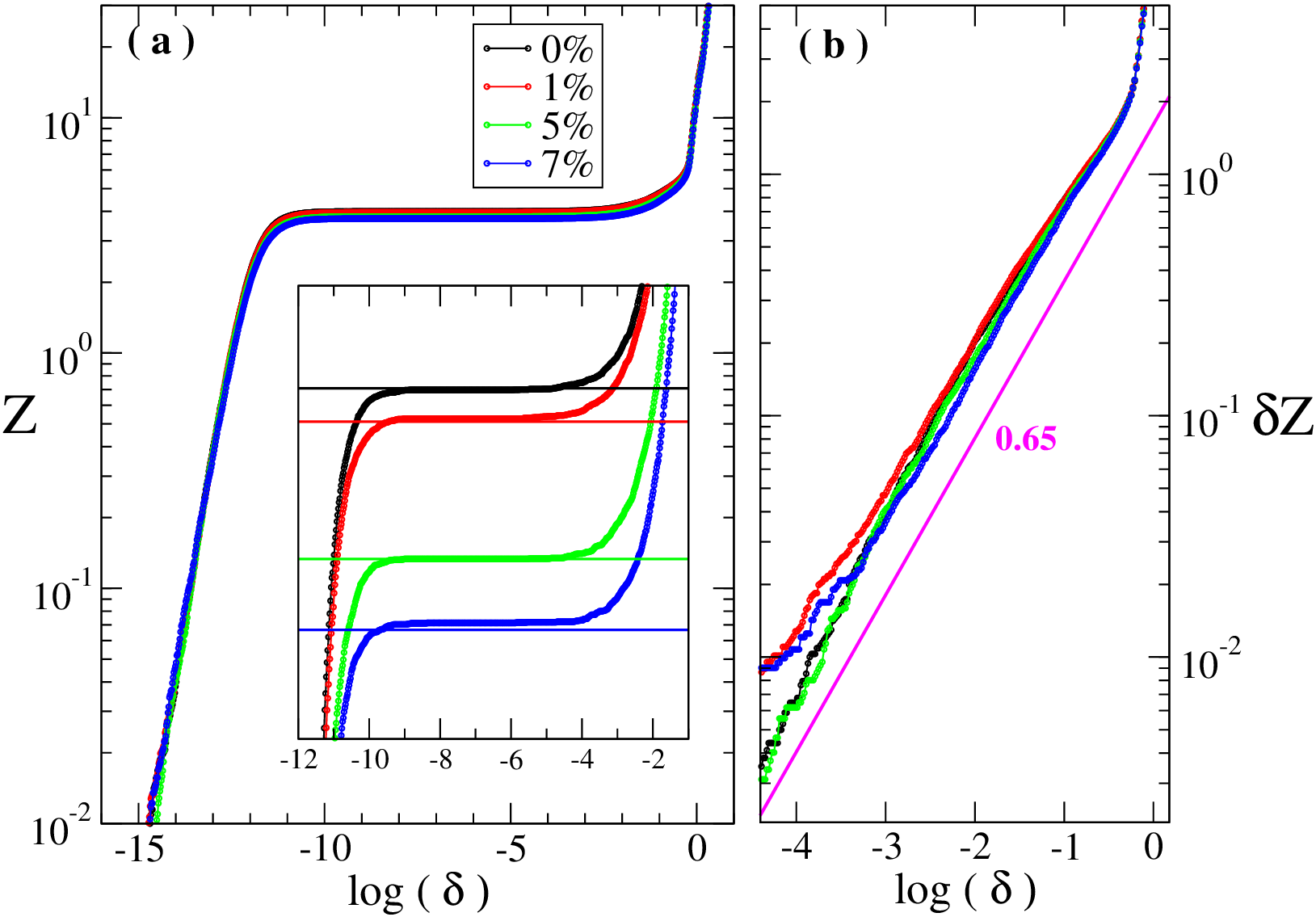}
\caption{
Integrated pair correlation $Z(\d)$~\cite{DTS05} for isostatic packings and several $f$, at pressure $p=10^{12}$.
Here rattlers have been removed (see text).
It shows a plateau at the isostatic value (a) and a power-law
growth beyond it (b). 
}
\label{IntGr}
\end{figure}

Fig.~\ref{IntGr} shows $Z(\d)$ for isostatic packings. Note that here ``rattlers'' (i.e. particles that do not participate to the jammed structure and rattle freely inside their
cage) have been removed following the procedure of~\cite{DTS05}: at the highest pressure $p=10^{12}$,
we identify as ``contacts'' the particles pairs that have gaps $\d \leq 10^{-8}$, and we then remove recursively
from the calculation of $Z(\d)$ those particles that
have strictly less than three contacts. Note also that because here particles are pinned at very low pressures, the gap between pairs of pinned particles is typically
very large so these contacts do not contribute to $Z(\d)$ at small $\d$.
We find that $Z(\d)$ has the same behavior as for unpinned jammed packings: it 
first grows on a scale $\sim 1/p =10^{-12}$ to a plateau that gives the average number of contacts~\cite{DTS05}.
The continuous line in the inset of the figure shows the theoretical prediction for the value of $z_J$, indicating that it is consistent 
with the generalized isostatic condition. 
Then, $Z(\d)$ grows as $\d^\a$ with $\a \sim 0.6$, which is also a signature of jamming~\cite{DTS05}.
We found that the exponent $\a$ is not affected by random pinning.

\begin{figure}
\centering
\includegraphics[width = 0.95\columnwidth, clip]{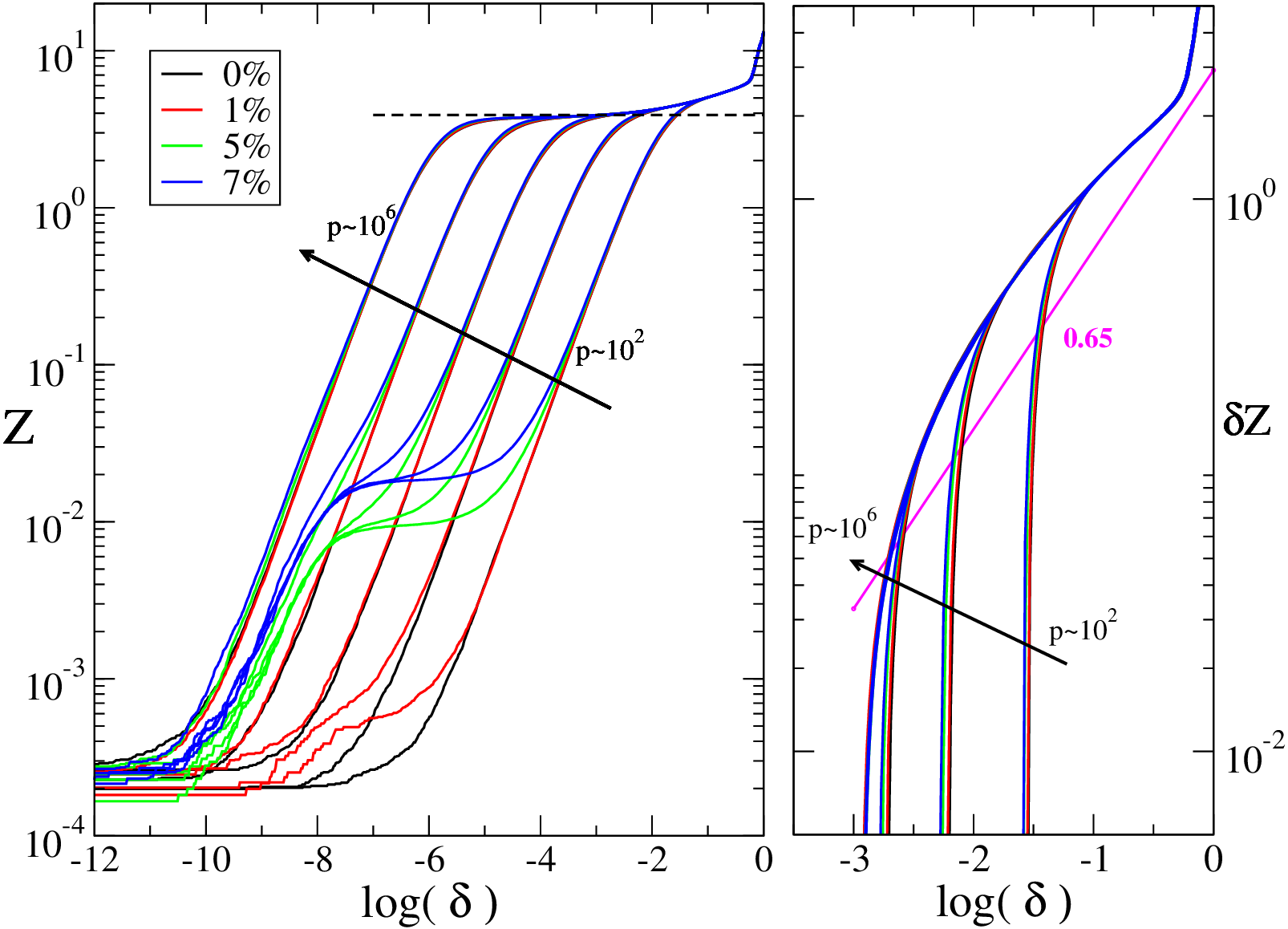}
\caption{
Integrated pair correlation $Z(\d)$~\cite{DTS05} for hyperstatic packings and several $f$.
The power-law growth beyond the plateau is more difficult to see (b). 
}
\label{IntGrHyp}
\end{figure}

Fig.~\ref{IntGrHyp} shows $Z(\d)$ for hyperstatic packings. Note that even at small pressures some
particles are found at $\d \sim 10^{-10}$: these are obviously pair of particles that have been pinned at high pressure and are therefore kept their
very small distances. Here, rattlers have not been removed because our focus is not on the isostatic condition
which we know to be violated. We focus instead on the growth of $Z(\d)$ above the plateau, at $\d \gg 1/p$. In this regime we found that 
$Z(\d)$ is independent of pinning. This is interesting because in absence of pinning, in this regime $Z(\d)$ grows roughly as $\d^{\a}$, as previously discussed. 
Although this power-law regime is barely visible in Fig.~\ref{IntGrHyp}(b), because of the uncertainty in the determination of the plateau
in presence of rattlers, the fact that $Z(\d)$ is independent of $f$ is enough to show that the power-law growth is not affected by pinning, even
for hyperstatic packings. This is interesting because the power-law growth of $Z(\d)$ has been connected to isostaticity~\cite{wyart_thesis}, while our data suggest
that this is not the case (although it is not clear that the argument of~\cite{wyart_thesis} can be applied in presence of random pinning). 
A better understanding of the relation between the power-law growth of $Z(\d)$ and isostaticity and marginality would be useful to clarify
the origin of this discrepancy.

\subsection{Mechanical stability condition and the microscopic structure} 

Independently of the protocol to generate jammed packings, it has been shown that, for $f=0$, 
many mechanical properties of the system are governed by the distance to the jamming $z_J$,  $\delta z=z-z_J$~\cite{PhysRevE.68.011306, wyart_thesis, Ellenbroek06,Wyart_EPL2005,Wyart05,BritoWyart_JCP2009}.

As it was explained above, the global mechanical stability criterion requires that the total number  of independent contact force components be bigger than the number of degrees of freedom for the packing to be mechanically stable~\cite{PhysRevE.60.687}. This  translates in a condition in the coordination number of a packing $z$, that has to be $z \ge 2d$  for a packing to be mechanically stable.
At the jamming point, it can be shown theoretically that $z=2d$ and it was measured numerically that this is the case \cite{PhysRevE.68.011306}.
This means that the packings at the jamming point are {\it marginally stable}: if any contact is broken, then the system is not rigid anymore~\cite{wyart_thesis}.

If the system is driven far from jamming -- even above or below $\phi_J$ -- it is under a finite compression and the criterion of rigidity is more demanding. It was derived for soft packings that, whenever a pressure $p$ appears in the system, the coordination number $z$ has to increase in order to keep rigidity in the packing  \cite{Wyart05}.  This  criterion was later extended for hard sphere system at finite pressure \cite{BritoWyart_JCP2009}. Packings of hard spheres at finite pressure have to respect the following relation in order to be rigid:
\beq
\delta z \ge A \sqrt{1 / p}.
\label{mech_crit}
\eeq
This relation has been tested numerically for systems with $f=0$ and it was observed that hard spheres obeys it {\it marginally}, i.e. with the equality,
in the entire glass phase~\cite{BritoWyart_EPL2006}, exactly like the jammed system does at $p\to\io$.

\begin{figure}
$\newline$
\centering
\includegraphics[width = 0.95\columnwidth, clip]{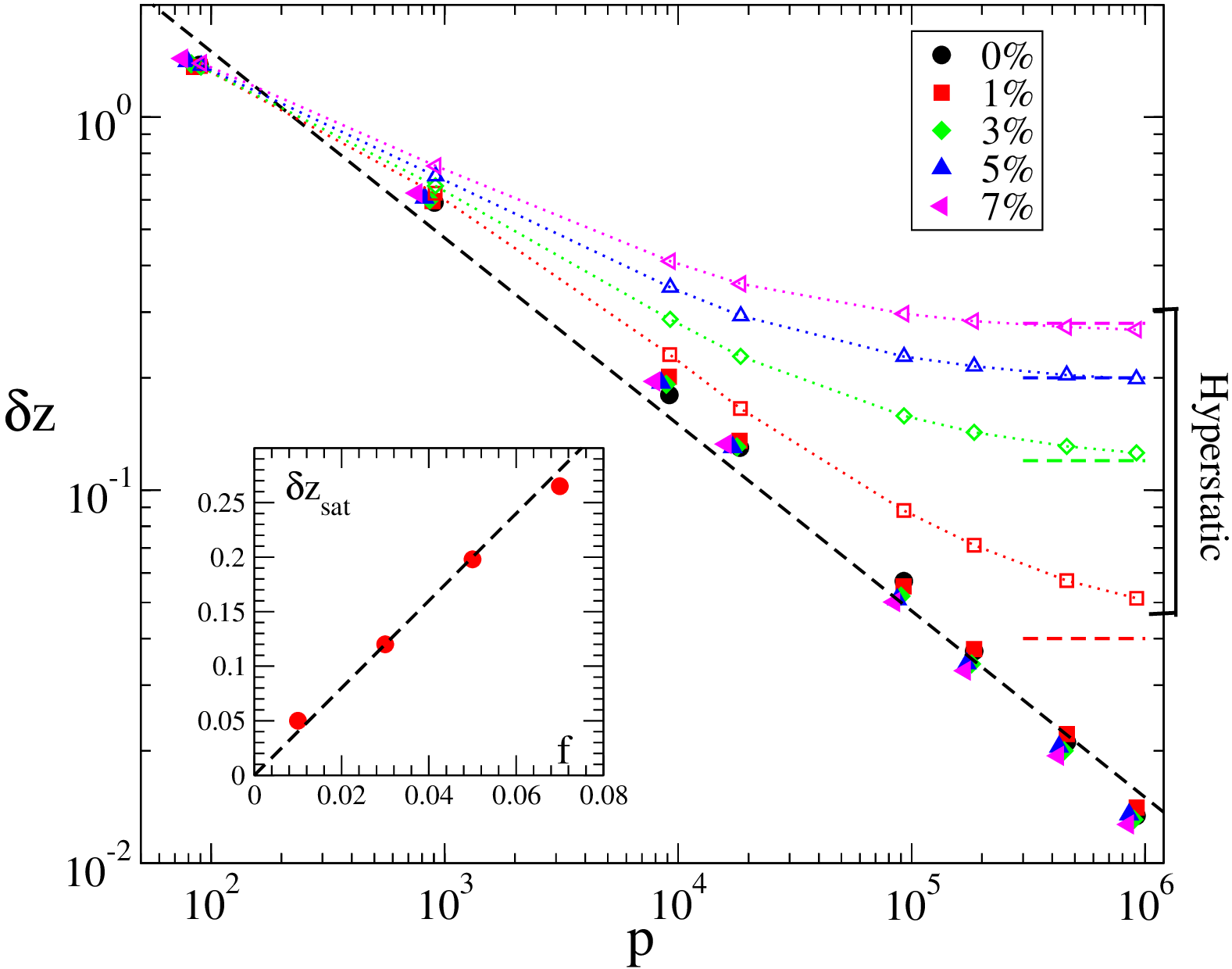}
\caption{$\delta z= z-z_J$ versus $p$ for different $f$, and both isostatic and hyperstatic packings.
The dashed line is the marginal stability in Eq.~(\ref{mech_crit}), $\delta z = A p^{-1/2}$.
Open symbols refer to hyperstatic initial configurations, and closed symbols 
correspond to isostatic initial configurations. Inset: $\delta z_{sat}$  versus $f$, where   
$\delta z_{sat}$ is the value of $\delta z$ measured at $p=10^6$, which is the maximum value of pressure simulated.
The dashed line corresponds to $\delta z = 2df$, the value of the coordination of hyperstatic packings at $p=10^{12}$, by construction.}
\label{dz_vs_p}
\end{figure}

We now test this relation for isostatic and hyperstatic packings generated using the protocols explained in section~\ref{methods}.
In Fig.~\ref{dz_vs_p} we plot a diagram of mechanical stability, where $\delta z$ is plotted versus $p$ for several $f$ and the 2 kinds of packings.
For packings that were isostatic at the jamming point, we observe that, as for $f=0$, the curves for $f > 0$ are compatible with the equality
 in Eq.~(\ref{mech_crit}) at all pressures, indicating that the system remains {\it marginally stable} in presence of random pinning.
In the case of the packings that were hyperstatic at the jamming point,  the system remains much more coordinated that is required to be rigid 
even when pressure is finite. These packings only approach the marginal stability line when pressure becomes relativelly small.  
We note that, by construction of our packings, at $p=10^{12}$, the coordination has to be  $z=2df$ (represented by the horizonal lines in the figure).
We also observe that the hexagonal crystals we considered have $\delta z\approx 2$ for any value of pressure, hence they are much more coordinated than any 
of the amorphous packings considered here.

This diagram of rigidity has important implications in the microscopic dynamics of the system, which are discussed in the next section.

\subsection{Microscopic dynamics}

\begin{figure}
$\newline$
\centering
\includegraphics[width = 0.95\columnwidth, clip]{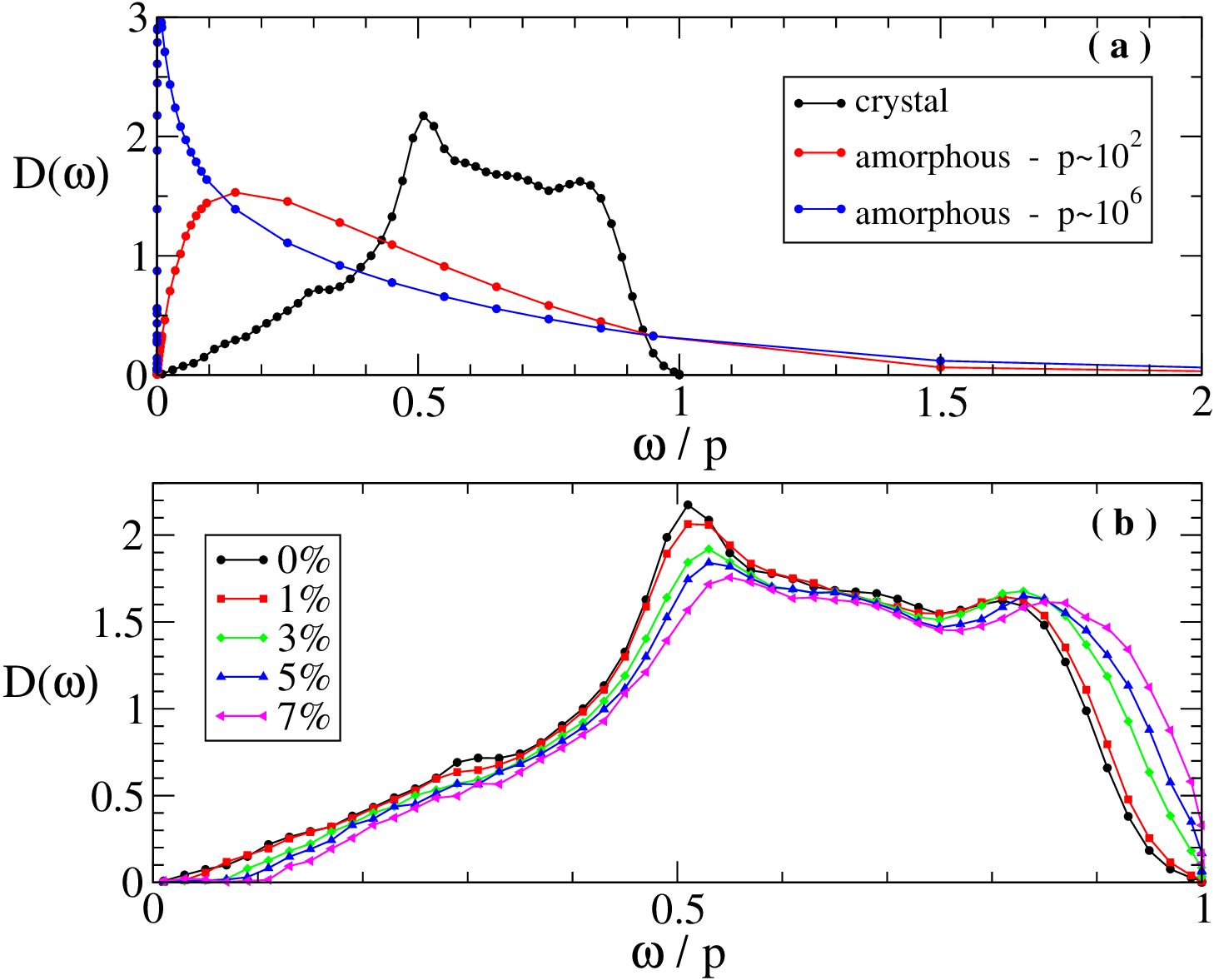}
\caption{{\bf (a)} Comparison of the density of states $D(\omega)$ of a crystal and glass at two different $p$ and $f=0$. 
{\bf (b)} $D(\omega)$ of a crystal with 5 different $f$, 
showing that the whole spectrum is shifted to higher frequencies when $f$ increases. 
All frequencies are rescaled by $p$ because the characteristic frequency is inversely proportional to $p$~\cite{BritoWyart_JCP2009}. 
%$N=4096$ for crystal and amorphous.
}
\label{Dw_crystal}
\end{figure}

\begin{figure}
$\newline$
\centering
\includegraphics[width = 0.95\columnwidth, clip]{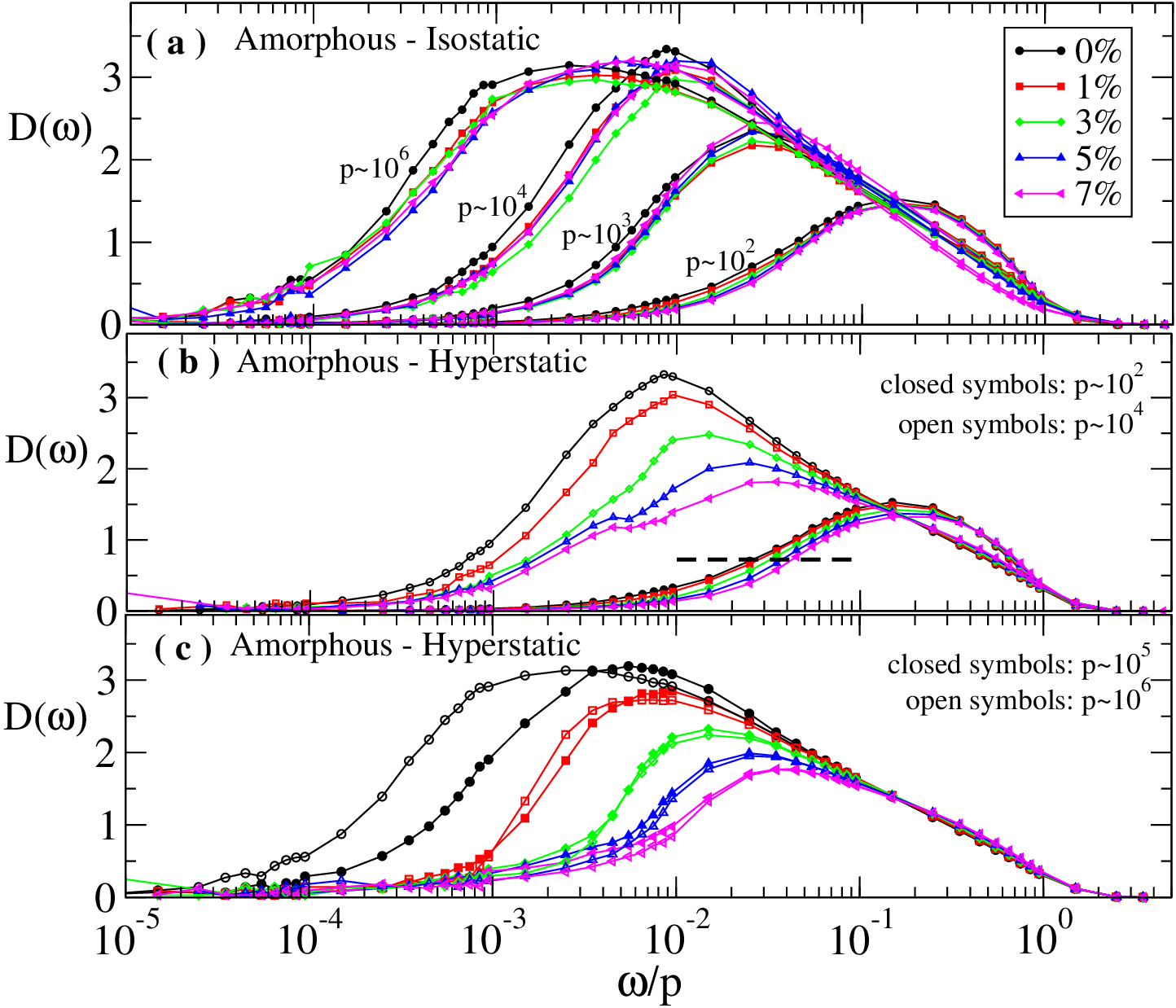}
\caption{$D(\omega)$ of a glass at different $f$ and different pressures  $p$ for
 {\bf (a)} isostatic initial configurations, 
 {\bf (b)} hyperstatic initial configurations at relatively small pressures and 
 {\bf (c)} hyperstatic initial configurations at high  pressures.}
\label{Dw_amorphous}
\end{figure}

%An important consequence of the fact that the system is marginally stable is the presence of {\it soft modes}~\cite{Wyart_EPL2005, BritoWyart_JCP2009}.
There is a direct relation between the mechanical stability of a packing and its vibrational modes: a rigid system cannot have unstable modes \cite{Ashcroft-Mermin}.
Then, if a system is just marginally rigid, this implies that it will have modes of very low frequency, which are on the verge of becoming unstable.
These {\it soft modes}~\cite{Wyart_EPL2005, BritoWyart_JCP2009} are revealed by measures of $D(\omega )$ shown in Fig.~\ref{Dw_crystal} and Fig.~\ref{Dw_amorphous}.
A comparison of a typical $D(\omega )$ in the crystal phase with two amorphous states at different pressures for $f=0$ is reported in Fig.~\ref{Dw_crystal}a.
The crystal displays the Debye behavior at small frequencies ($D(\omega) \sim \omega$) while the amorphous configurations present a strong excess of low-frequency 
modes.
For the crystal, as soon as a fraction $f$ of particles is pinned, the whole spectrum is shifted to higher frequencies, Fig.~\ref{Dw_crystal}b, 
indicating that plane-waves with the smallest frequencies disappear, like the softest standing wave in a string that is pinned at half of its length.

For the amorphous system with isostatic initial conditions, one observes that the pinning does not affect the spectrum strongly, 
as can be seen in Fig.~\ref{Dw_amorphous}a for different values of $f$ and different pressures. 
Note that for all pressures the spectrum is roughly the same for all values of $f$ we studied. 
However, in the amorphous case with hyperstatic initial conditions, the result is different depending on the pressure.
Fig.~\ref{Dw_amorphous}b shows the spectrum of vibrations for relatively small pressures, $p\approx 10^2$ and $p\approx 10^4$: 
in this range of pressure, pinning has a small effect on the vibrations.
The picture is different in  the case of high pressure, shown in the Fig.~\ref{Dw_amorphous}c.
In this case the curves at the same value of $f$ and different pressures are the same, and pinning has a strong effect and suppresses the low frequency
part of the spectrum.
This means that, above a certain pressure, the modes of very low frequency are suppressed.

\begin{figure}
$\newline$
\centering
\includegraphics[width = 0.97\columnwidth, clip]{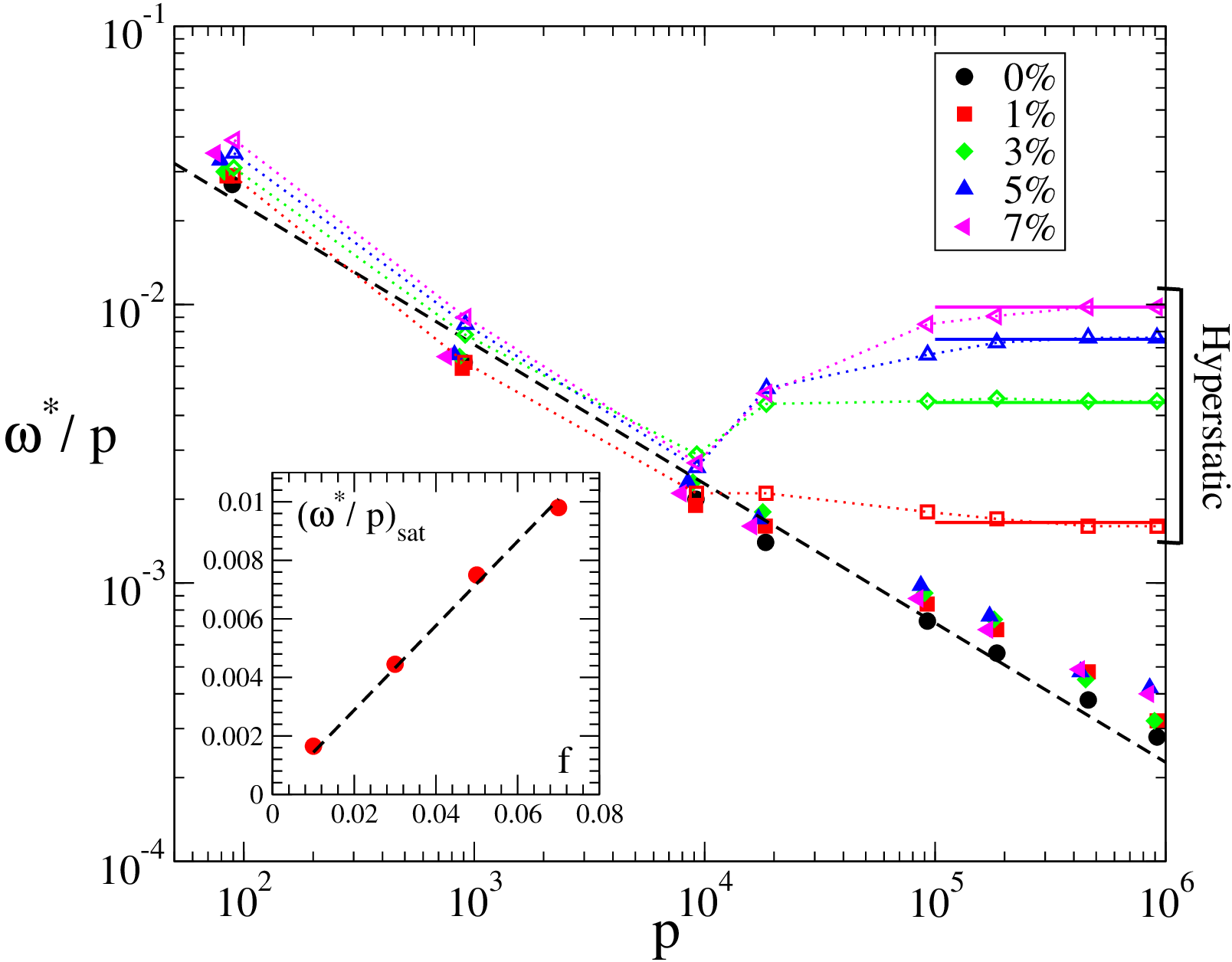}
\caption{ 
Plot of $\omega^*/p$ versus $p$, for several $f$ and both isostatic and hyperstatic packings. 
The black dashed line corresponds to $\omega^*/p \sim p^{-1/2}$.
Open symbols refer to hyperstatic initial  configurations and closed symbols refer 
to isostatic initial configurations.
Inset: $(\omega^*/p)_{sat}$  versus $f$. The black dashed line corresponds to $(\omega^*/p)_{sat} \sim 2df$.}
\label{w_vs_p}
\end{figure}

The soft modes of the glass have a characteristic frequency $\omega^*$, and if marginal stability in Eq.~(\ref{mech_crit}) is 
assumed throughout the glass phase, one expects that $\omega^*/p \sim p^{-1/2}$~\cite{BritoWyart_JCP2009}. 
We define $\omega^*$ as the value such that $D(\omega)$ is equal to half of its maximum (as marked by a horizontal dashed line 
in Fig.~\ref{Dw_amorphous}b to exemplify in the case for the smallest pressure), and we plot this $\omega^*$ in Fig.~\ref{w_vs_p}.
In the case of the amorphous system with isostatic initial conditions, for all values of $f$, we observe a nice agreement between 
the data and the theoretical curve, evidencing the marginal character of the glassy phase and indicating that the presence of some 
frozen particles in the system does not prevent the development of low-frequency modes.
For the amorphous case with hyperstatic initial conditions one observes that $\omega^*/p$ follows the theoretical line until 
$p\approx 10^4$, when the characteristic frequency saturates. 
In the inset of Fig.~\ref{w_vs_p}, we plot the frequency at which the curve saturates, named $(\omega^*/p)_{sat}$  as a function of $f$.
The  dashed line is a fit of the theoretical prediction $\omega^*/p \sim \d z \sim 2df$  ~\cite{Wyart_EPL2005,wyart_thesis}, which 
is in a very nice agreement with our simulations.

%\begin{figure}
%$\newline$
%\centering
%%\includegraphics[width = 0.95\columnwidth, clip]{figs_new/graf_Dw-IsoHyperCrystal.eps}
%\includegraphics[width = 0.95\columnwidth, clip]{graf_Dw-IsoHyperCrystal.eps}
%\caption{\textcolor{red}{alternative figure: it would replace figs \ref{Dw_crystal} and \ref{Dw_amorphous} }}
%\label{xx}
%\end{figure}
%

\section {Discussion and Conclusion}

This work has three main conclusions:
(i) pinned particles do not change the jamming transition provided that configurations are isostatic, in the generalized sense defined above; 
(ii) the low frequency modes in a crystal behave differently from the modes of the amorphous solid under pinning; and 
(iii) the only relevant parameter in the system is the distance to the isostatic point.
We now discuss each of these conclusions in more detail.

The presence of pinned particles is taken into account by considering that some degrees of freedom are frozen,  leading
to a generalization to the concept of isostaticity,  $z_J=2d(1-f)$. 
Using the protocol in which we pin particles while the system is still at equilibrium, we generated configurations that numerically satisfy 
 to extremely good accuracy  the condition of isostaticity at the jamming point, $p=10^{12}$ (Fig.~\ref{IntGr}).
When these isostatic configurations are used as initial conditions to study the glass phase at smaller pressure, they generate configurations that 
are marginally stable, with $\d z = z - z_J = A \sqrt{1/p}$, which is also very well satisfied by numerical data (Fig.~\ref{dz_vs_p}). 
 This marginal stability condition implies that the system has an excess of low frequency modes, with soft modes characterized by a characteristic
frequency $\omega^*/p \sim \d z$ ~\cite{Wyart_EPL2005,wyart_thesis}, which are not affected by pinning (Fig.~\ref{w_vs_p}).
 Moreover, near contacts are characterized by a power-law growth of $Z(\d)$ that is not affected by the pinning of particles as well.
All this together allows us to conclude that the jamming transition remains qualitatively similar in the presence of randomly pinned particles provided that the 
notion of generalized isostaticity holds.

Given that pinning of some particles breaks the translational symmetry, the fact that  pinning does not affect the properties of jamming implies that 
translational symmetry does not play a role in the properties of nearly jammed systems.
Note that this represents a deep difference between crystals and amorphous system in what concerns the {\it nature} of their vibrational modes: in crystals, 
plane waves are a consequence of the (spontaneously broken) translational symmetry and then the pinning of particles does change their modes, shifting them to higher frequencies 
(Fig.~\ref{Dw_crystal}).

In the case where hyperstatic configurations are used to study the glass phase, simulations show that, even when the system is away from the jamming point, it 
keeps being overcontrained (Fig~\ref{dz_vs_p}). Then, since the system has more coordination than it is necessary to keep rigidity, the soft modes of low frequency are
stabilized: the modes with zero frequency are shifted to higher frequency (Fig~\ref{w_vs_p}). 
The excess of coordination with respect to the isostatic value is $\delta z=2df$ and this implies that $\omega^*/p \sim \d z \sim 2df$, 
which is in agreement with our simulations (inset of Fig~\ref{w_vs_p}).

It has been shown that the characteristic frequency of the modes is related to a length scale that is inversely proportional to the distance to the 
isostatic point $l^* \sim p/\omega^* \sim 1/\d z$ ~\cite{Wyart_EPL2005,wyart_thesis}. 
Because our data are in agreement with all the predictions derived from the existence of this length scale,
this suggests that $l^*$  is the only important length scale in the problem: this length seems to dominate the physics of vibrations in the amorphous system. 
Note that the pinning of some particles introduces a new length that is the average distance between randomly distributed fixed particles, $l_f \sim f^{-1/d}$.
Then, one could expect that for a certain value of $f>0$ and at big values of $p$,  $l^*$ is so big that  $l^*>l_f$  and then the softest modes of 
the system would be stabilized because of that. However, this is not observed: when considering isostatic configurations, the presence of 
pinning does not change the  frequency of the natural vibrations.
% and in the case where we use hyperstatic configurations,  the natural 
%frequency of the modes does change, but it is well captured by $l^*$.
Then it seems that $l^f$ is really not important  for the vibrations.
There are two possible explanations for this fact. 
First, it might be that the soft modes change in nature in the presence of pinning in order to adapt to the pinned geometry while keeping their low energy. 
This is possible due to their amorphous nature.
The other possibility is that our simulations are always in a limit of ``weak pinning'' \cite{Cammarota2011} and $l^* < l_f$ always.
Given however that we explored quite large pinning fractions and very high pressures, this second possibility seems less likely, and we therefore
conclude that pinning particles and the associated breaking of translational invariance does not have a strong impact on the physics of the jamming
transition if the configurations are isostatic.

\paragraph*{Acknowledgements --} We thank Giulio Biroli and Marco Tarzia for useful 
contributions at the early stage of this project,
and Matthieu Wyart for useful comments.
Financial support was provided by
the European Research Council through ERC grant agreement
no. 247328 and by CAPES/Cofecub through Grant No. 667/10.

\bibliography{modes_references}

\end{document}